\documentclass[aps,10pt,pre,byrevtex,twocolumn,superscriptaddress,bibnotes,longbibliography]{revtex4-1}
\usepackage{microtype}
\usepackage{graphicx}
\usepackage{amsmath}
\usepackage{amssymb}
\usepackage{color}
\definecolor{myblue}{rgb}{0.153,0.322,0.706}
\usepackage[colorlinks,linkcolor=myblue,urlcolor=myblue,citecolor=myblue]{hyperref}

\newcommand{\be}{\begin{equation}}
\newcommand{\ee}{\end{equation}}
\newcommand{\idf}{1\!\! 1}
\newcommand{\ra}{\rightarrow}
\newcommand{\cL}{\mathcal{L}}
\newcommand{\cH}{\mathcal{H}}
\newcommand{\reals}{\mathbb{R}}
\newcommand{\conf}{\text{conf}}
\newcommand{\esc}{\text{esc}}
\newcommand{\hX}{\hat X}
\newcommand{\eps}{\varepsilon}

\begin{document}

\title{Dynamical phase transition in drifted Brownian motion}

\author{Pelerine Tsobgni Nyawo}
\email{tsobgnipelerine@gmail.com}
\affiliation{Department of Physics, Stellenbosch University, Stellenbosch 7600, South Africa}

\author{Hugo Touchette}
\email{htouchet@alum.mit.edu, htouchette@sun.ac.za}
\affiliation{\mbox{Department of Mathematical Sciences, Stellenbosch University, Stellenbosch 7600, South Africa}}
\affiliation{National Institute for Theoretical Physics (NITheP), Stellenbosch 7600, South Africa}

\date{\today}

\begin{abstract}
We study the occupation fluctuations of drifted Brownian motion in a closed interval, and show that they undergo a dynamical phase transition in the long-time limit without an additional low-noise limit. This phase transition is similar to wetting and depinning transitions, and arises here as a switching between paths of the random motion leading to different occupations. For low occupations, the motion essentially stays in the interval for some fraction of time before escaping, while for high occupations the motion is confined in an ergodic way in the interval. This is confirmed by studying a confined version of the model, which points to a further link between the dynamical phase transition and quantum phase transitions. Other variations of the model, including the geometric Brownian motion used in finance, are considered to discuss the role of recurrent and transient motion in dynamical phase transitions.
\end{abstract}

\maketitle

\section{Introduction}

We continue in this paper our study of the occupation fluctuations of drifted Brownian motion (dBM) \cite{tsobgni2016b}. The motivation for studying this model is that it shows a dynamical phase transition (DPT), that is, a sudden change in the way that fluctuations are created in the long-time limit, leading to singularities in large deviation functions, the nonequilibrium analogs of thermodynamic potentials \cite{touchette2009}. Similar DPTs are found in interacting particle systems such as kinetically constrained models of glasses \cite{garrahan2007,garrahan2009,jack2014} and the exclusion process \cite{bodineau2005,espigares2013,hurtado2014,tizon2016,lazarescu2017}, which show DPTs in the integrated activity or current for some parameter values. In these and many other models, however, a DPT arises when taking the long-time limit in addition to a hydrodynamic or macroscopic limit \cite{derrida2002a,bunin2013,baek2017,shpielberg2017}, which is equivalent to a low-noise limit \cite{freidlin1984,graham1989,bertini2015b}. 

The advantage of dBM is that its DPT arises in the long-time limit without a low-noise limit, making it an ideal model to investigate general or minimal conditions for the appearance of DPTs. It is known, for instance, that DPTs cannot arise without a low-noise limit in ergodic Markov processes evolving on finite or compact spaces \footnote{See Theorem 3.1.2 of \cite{dembo1998} for the case of finite Markov chains. The case of compact diffusions with additive observables follows by contraction of the ``level-2'' results of G{\"a}rtner \cite{gartner1977}, generalized to ``current'' observables in \cite{bierkens2013}.}; yet it is not clear what properties of unbounded processes, such as Langevin-type diffusions in $\reals^d$, are responsible for the appearance of DPTs. The ``unboundedness'' of the state space is certainly not sufficient, which means that other properties such as ergodicity, confinement or recurrence might play a role. Recently, it has been found that the large deviations of non-homogeneous random walks with resetting can also have DPTs in the long-time limit \cite{majumdar2015b,harris2016}, bringing new questions about the relation between time-dependent driving and DPTs \footnote{The DPT considered in \cite{majumdar2015b} is not related to large deviations as such, but to different scalings of the time-dependent distribution in the stationary limit.}.

Here, we focus on the role of confinement and recurrence in diffusions by showing that the DPT found in the occupation large deviations of dBM is related to a confinement-escape transition in the atypical paths of this model. This transition is similar to first-order DPTs arising in processes with absorbing states, and can also be seen as a dynamical or fluctuation analog of wetting and depinning transitions. What drives the escape transition in dBM is the fact that it is not recurrent when it has a drift. This is confirmed by considering a confined version of the model and by studying its large deviations in the null confinement limit. With this model, we also establish an interesting connection between DPTs and quantum phase transitions \cite{sachdev2011}.

We define in the next sections the dBM model and present a complete account of its occupation large deviations and of its DPT, first announced in \cite{tsobgni2016b}, which is fundamentally related to the non-Hermitian nature of the spectral problem underlying long-time large deviations \cite{touchette2017}. We complement these results by studying in detail the so-called driven or auxiliary process \cite{jack2010b,chetrite2013,chetrite2014,chetrite2015}, which explains in our case how fluctuations of the occupation are created in the long-time limit \cite{angeletti2015}, and by presenting simulation results that confirm the confinement and escape regimes. We close by discussing other models based on Brownian motion, including the geometric Brownian motion, for which escape or deconfinement DPTs are also expected to arise. 
	
\section{Model}

We consider a dBM on $\reals$ \cite{tsobgni2016b}, defined by the stochastic differential equation (SDE)
\be
dX_t = \mu dt+\sigma dW_t
\ee
or, equivalently, by its solution
\be
X_t = \mu t +\sigma W_t
\ee 
with $X_0=0$. Here, $\mu$ is the drift, $W_t$ is the standard Brownian motion (BM) on $\reals$ with $W_0=0$, acting as a noise, and $\sigma>0$ is the noise amplitude. This model represents in the simplest case a particle moving at constant velocity $\mu$, perturbed by a Gaussian white noise originating from thermal noise or background vibrations \cite{goohpattader2009}. The variable $X_t$ can also be interpreted as the log-return of a stock price with mean $\mu$ and volatility $\sigma$ \cite{bouchaud2000b}, or as the random charge dissipated in a resistor when applying a linearly-increasing voltage in time, in which case $W_t$ is a Nyquist noise \cite{garnier2005}.

For a given time interval $[0,T]$, we study the fluctuations of the time that $X_t$ spends in some subset $A\subset\reals$, as expressed by the integral
\be
R_T =\int_0^T \idf_{A}(X_t)\, dt,
\ee
where $\idf_A(x)$ is the characteristic or indicator function of $A$, equal to $1$ if $x\in A$ and $0$ otherwise. This residence or occupation time has been studied extensively in probability theory \cite{darling1957,levy1965,chan1994,gruet1996,rouault2002,mansuy2008} and physics \cite{berezhkovskii1998,godreche2001,majumdar2002b,majumdar2002c,barkai2006,burov2007}, starting with L\'evy \cite{levy1940} who derived his well-known arcsine law for the occupation time of BM in $A=[0,\infty)$, generalized to dBM by Akahori \cite{akahori1995} and Dassios \cite{dassios1995}. Here, we take $A$ to be a closed interval $[a,b]$ and consider the \emph{occupation fraction} $\rho_T = R_T/T$ so as to obtain a random variable taking values on $[0,1]$ with a probability density that scales according to
\be
P(\rho_T=\rho)\approx e^{-T I(\rho)}
\ee
for large times $T\gg 1$. This approximation is called the \emph{large deviation principle} (LDP) and implies that $P(\rho_T=\rho)$ decays exponentially in $T$, at leading order in $T$, with a decay rate controlled by the function $I(\rho)$, called the \emph{rate function} \cite{ellis1985,dembo1998,hollander2000}. This function is positive and is equal to $0$ here only for $\rho=0$, which means that $\rho_T\ra 0$ with probability 1 as $T\ra\infty$. This only translates the fact that dBM has no stationary distribution (or, formally speaking, a flat invariant distribution), so that it is more likely to stay outside than inside the interval $[a,b]$ for long times. The LDP and its rate function characterizes the exponentially small probability that the dBM visits that interval for a fraction $\rho$  of time.

The method that we use to calculate the rate function is based on the G\"artner--Ellis Theorem \cite{ellis1985,dembo1998,hollander2000}, which gives $I(\rho)$ as the Legendre--Fenchel transform of the \emph{scaled cumulant generating function} (SCGF),
\be
\lambda(k)=\lim_{T\ra\infty} \frac{1}{T}\ln \langle e^{Tk\rho_T}\rangle,
\ee
provided that this function is differentiable as a function of the real parameter $k$ conjugated to $\rho_T$. Under this condition, we thus have
\be
I(\rho) = \sup_{k}\{k \rho-\lambda(k)\}.
\label{eqlf1}
\ee

To find $\lambda(k)$, we then use the fact that the generating function $\langle e^{Tk\rho_T}\rangle$ evolves linearly with $T$, which leads us to express the limiting function $\lambda(k)$ as the principal eigenvalue of some linear operator, corresponding to the generator of that evolution \cite{touchette2017}. In our case, this generator is a linear differential operator, given by
\be
\cL_k = L+k\idf_{[a,b]}(x),
\label{eqgen1}
\ee
where
\be
L=\mu\frac{d}{dx}+\frac{\sigma^{2}}{2}\frac{d^2}{dx^2}
\label{eqgen2}
\ee
is the Markov generator of dBM. The spectral problem that we need to solve to obtain the SCGF is therefore
\be
\cL_k r_k(x) =\lambda(k) r_k(x),
\label{eqeig1}
\ee
where $\lambda(k)$ is the principal eigenvalue of $\cL_k$ and $r_k>0$ is its corresponding eigenfunction. The boundary conditions on $\reals$ that must be used to solve this problem are as follows \cite{touchette2017}. Because $\cL_k$ is non-Hermitian, one must consider the dual problem
\be
\cL_k^\dag l_k=\lambda(k) l_k,
\label{eqeig2}
\ee
where $\cL_k^\dag$ is the dual of $\cL_k$ with respect to the standard (Lebesgue) scalar product, and impose that the product $r_k(x)l_k(x)$, which is positive, decay sufficiently fast to $0$ as $x\ra\pm\infty$ to be integrable. The normalization is then set by
\be
\int_{-\infty}^\infty r_k(x)\, l_k(x)\, dx =1.
\label{eqnorm1}
\ee

The above method for calculating large deviation functions is standard \cite{touchette2009,touchette2017,angeletti2015,majumdar2002}. Another method based on the ``level 2'' of large deviations is described in Appendix A.2 of \cite{chetrite2015} or in \cite{angeletti2015}. The two methods are equivalent, in that it can be shown that the solution of the level-2 method is the product function
\be
p_k(x) = r_k(x)l_k(x),
\ee
which has the interpretation of a probability density. As explained in \cite{angeletti2015}, this is the stationary density of a modification of the process $X_t$, called the \emph{driven process}, interpreted as the process describing the subset of trajectories of $X_t$ leading to a given fluctuation $\rho_T=\rho$ \footnote{Pinsky \cite{pinsky1985} considers the special case $\rho=1$ where $X_t$ is conditioned to always stay in some interval.}. 

We refer to \cite{jack2010b,chetrite2013,chetrite2014,chetrite2015} for more information about the construction and interpretation of this process, also known as the \emph{auxiliary}, \emph{fluctuation} or \emph{conditioned process} \cite{jack2010b}. For our purpose, note that the driven process associated with the occupation fluctuations of dBM is the diffusion $\hX_t$ satisfying the new SDE,
\be
d\hX_t =F_k(\hX_t) dt+\sigma dW_t,
\label{eqdp1}
\ee
where 
\be
F_k(x) = \mu+\sigma^2\frac{r_k'(x)}{r_k(x)}
\label{eqeff1}
\ee
is a space-dependent drift or force that modifies the constant drift of dBM. Moreover, choosing $k$ such that 
\be
\lambda'(k)=\rho
\label{eqlf2}
\ee
leads $\hX_t$ to realize $\rho_T=\rho$ as a typical (ergodic) occupation, so we effectively transform with Eqs.~\eqref{eqdp1}-\eqref{eqlf2} what is an \emph{atypical} occupation for dBM into a \emph{typical} occupation for the driven diffusion \cite{angeletti2015}. In this sense, the driven diffusion provides a physical way to understand how occupation fluctuations are created by means of a modified force $F_k$, which is an effective or entropic force capturing the effect of the noise. This can be made more precise by showing that the driven process is equivalent to the process obtained by conditioning $X_t$ on reaching the occupation $\rho_T=\rho$ \cite{chetrite2014}, which gives a nonequilibrium version of the microcanonical ensemble in which only trajectories with that occupation are considered \cite{chetrite2013}.


For dBM, it is important to note that the constraints above on $r_k$ and $l_k$ cannot always be satisfied, as we will see next, because the model is non-confined. In particular, the left eigenvector $l_0$ obtained for $k=0$, which is the solution of the time-independent Fokker--Planck equation,
\be
\cL_0^{\dag} l_0(x) = -\mu l'_0(x) +\frac{\sigma^2}{2}l''_0(x) = 0,
\label{eqfp1}
\ee
cannot be normalized on $\reals$, and neither can $r_0l_0$. In this case, $r_0$ must be constant in order to consistently have $\lambda(0)=0$ and $F_0=\mu$ for $k=0$. This is important for understanding the DPT.

\section{Large deviations}

We solve in this section the spectral problem described before to obtain the SCGF, the rate function, and the driven diffusion characterizing the large deviations of the occupation fraction. The full solution of the problem involves two types of solutions that we discuss separately. Without loss of generality, we take $\mu\geq 0$ and consider the occupation interval $[-a,a]$ centered around $x=0$. Negative drifts and non-centered (closed) intervals can be treated by reflecting and translating $X_t$ properly.

\subsection{Quantum solution}

\begin{figure}[t]
\centering
\includegraphics{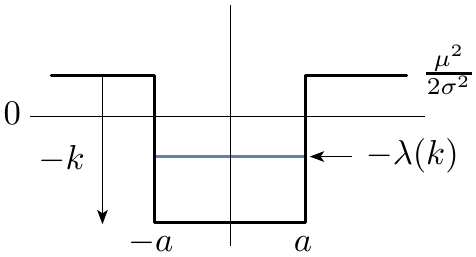}
\caption{Equivalent quantum well problem.}
\label{figqp1}
\end{figure}

The direct spectral problem \eqref{eqeig1} involving the \emph{tilted generator} $\cL_k$ is not Hermitian because of the first derivative term appearing in the generator \eqref{eqgen2} of dBM. However, it can be mapped to a Hermitian operator $\cH_k$ determining the spectral problem
\be
\cH_k \psi_k = \lambda(k) \psi_k
\ee
by applying a so-called \emph{symmetrization} to $\cL_k$ \cite{touchette2017}, defined by
\be
\psi_k = p_0^{1/2}r_k = p_0^{-1/2} l_k
\label{eqtransf1}
\ee
and
\be
\cH_k = p_0^{1/2} \cL_k p_0^{-1/2},
\label{eqsym1}
\ee
where $p_0 = r_0l_0 = l_0$ is normally the stationary distribution of the process considered. Here, there is no stationary distribution, but the symmetrization can nevertheless be applied with $l_0 = e^{2\mu x/\sigma^2}$, which solves Eq.~\eqref{eqfp1}, as a purely mathematical trick to remove the non-Hermitian term in $\cL_k$ and obtain 
\be
\cH_k = \frac{\sigma^2}{2}\frac{d^2}{dx^2}-V_k(x),
\ee
where
\be
V_k(x) = \frac{\mu^2}{2\sigma^2} -k \idf_{[-a,a]}(x).
\ee
Up to a minus sign, this is the Schr\"odinger equation for a finite square well, leading us to associate $\lambda(k)$, the top eigenvalue of $\cL_k$, to minus the ground-state energy of the well, as illustrated in Fig.~\ref{figqp1}. The boundary conditions for $r_k$ and $l_k$ translate for $\psi_k$ into normal quantum (Dirichlet) boundary conditions, namely, $\psi_k(x)\ra 0$ as $x\ra\pm\infty$, so that
\be
\int_{-\infty}^\infty \psi_k(x)^2\, dx=1,
\ee
which is the normalization in \eqref{eqnorm1} with \eqref{eqtransf1}.


\begin{figure}[t]
\centering
\includegraphics{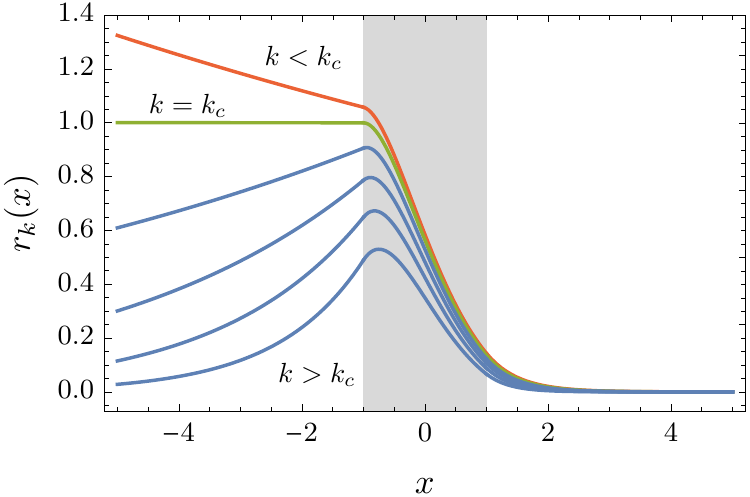}
\caption{(Color online) Quantum eigenfunction for different values of $k$. Parameters: $\mu=1$, $\sigma=1$, and $a=1$. The occupation region $[-a,a]$ is shaded in grey. }
\label{figqrk1}
\end{figure}

The solution of this quantum problem can be found in any quantum physics textbook. There is a bound ground state $\psi_k(x)$ for any well depth $-k<0$, made of a cosine in the well connected to two decaying exponentials on either side of the well. The corresponding eigenvalue is given by
\be
\lambda_q(k)=\lambda^0_q(k) -\frac{\mu^2}{2\sigma^2},
\label{eqqeig1}
\ee
$-\lambda_q^0(k)$ being the lowest eigenvalue of the non-raised well ($\mu=0$) solving the transcendental equation
\be
\zeta = \gamma\tan(\gamma a),
\ee
where
\be
\zeta =\frac{\sqrt{2\lambda_q^0}}{\sigma},\qquad \gamma =\frac{\sqrt{2(k-\lambda_q^0)}}{\sigma}.
\label{eqzeta1}
\ee
For $k=0$, we obviously have $\lambda_q^0(0)=0$ and therefore $\lambda_q(0)=-\mu^2/(2\sigma^2)$. For $k>0$, $\lambda_q(k)$ then increases monotonically from this negative value to become positive beyond a critical value of $k$, denoted by $k_c$, which depends on $\mu$ and $\sigma$. 

We show in Fig.~\ref{figqrk1} the corresponding ``right'' eigenfunction $r_k(x)$, given by \eqref{eqtransf1}:
\be
r_k(x) = e^{-\mu x/\sigma^2}\psi_k(x)
=
\left\{
\begin{array}{lll}
e^{(\zeta -\mu/\sigma^2)x} & & x<-a\\
A e^{-\mu x/\sigma^2}\cos(\gamma x) & & x\in [-a,a]\\
B e^{-(\zeta+\mu/\sigma^2)x} & & x>a,
\end{array}
\right.
\label{eqqsol1}
\ee
where $A$ and $B$ are constants fixed by imposing the continuity of $\psi_k$ or $r_k$ \footnote{The normalization of $\psi_k$ or $r_k$ is not required for the driven process, so we do not impose it.}. We can see that $r_k(x)$ decays to $0$ as $x\ra\infty$ because $\zeta+\mu/\sigma^2>0$ for all $k\geq 0$, but that it decays to $0$ as $x\ra-\infty$ only when $k>k_c$ because then $\lambda_q(k)>0$, so that  $\zeta-\mu/\sigma^2>0$. At the critical value $k=k_c$, $r_k(x)$ is constant for $x<-a$, with a height arbitrarily fixed at $1$.

We will analyse the driven process associated with this eigenfunction shortly. For now, note that $\psi_k(x)$ does not depend on $\mu$, as is obvious from the quantum problem (the wavefunction is invariant under vertical and horizontal translation of the well), which means that $r_k(x)$ depends on that parameter only via the symmetrizing factor $e^{-\mu x/\sigma^2}$, which makes $r_k(x)$ non-symmetric around $x=0$, compared to $\psi_k(x)$ which is symmetric. Moreover, the quantum eigenvalue $\lambda_q(k)$ depends on $\mu$ only via a trivial shift of the $\mu=0$ eigenvalue, as shown in (\ref{eqqeig1}).

\subsection{Non-quantum solution}

\begin{figure}[t]
\centering
\includegraphics{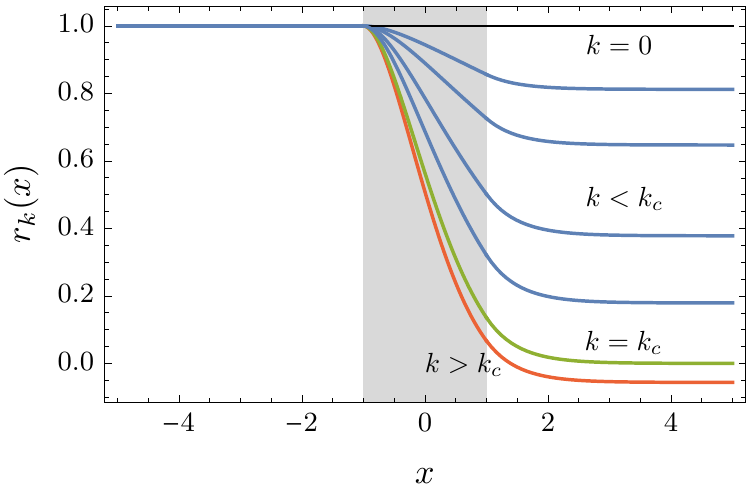}
\caption{(Color online) Non-quantum eigenfunction for different values of $k$. Parameters: $\mu=1$, $\sigma=1$, and $a=1$.}
\label{fignqrk1}
\end{figure}

The quantum solution obtained before cannot represent the whole SCGF because it does not satisfy $\lambda(0)=0$, while $r_k(x)$ does not converge to a constant function as $k\ra 0$. Based on the latter property, we now look for real continuous solutions of the non-Hermitian spectral problem \eqref{eqeig1} of the form
\be
r_k(x) =
\left\{
\begin{array}{lll}
1 & & x<-a\\
e^{-\mu x/\sigma^2}(A e^{-i\gamma' x}+Be^{i\gamma' x}) & & x\in[-a,a]\\
C e^{-2\mu x/\sigma^2} +D & & x>a,
\end{array}
\right.
\ee
where
\be
\gamma'=\frac{\sqrt{2k-\mu^2/\sigma^2}}{\sigma}
\ee
and $A$, $B$, $C$ and $D$ are constants fixed again to ensure continuity. It can be checked that there are non-trivial solutions for these coefficients all associated, remarkably, with the eigenvalue $\lambda_{nq}(k)=0$ for all $k\geq 0$, which we refer to as the ``non-quantum'' eigenvalue. This is obvious for the branches $x<-a$ and $x>a$, but can also be verified for $x\in[-a,a]$ by applying $\cL_k$ on this branch.

This solution for $r_k$ is plotted in Fig.~\ref{fignqrk1} for various values of $k$ above and below the critical value $k_c$. We can see that $r_k(x)=1$ when $k=0$, which is the correct eigenfunction associated with $\cL_{k=0}=L$. For $k>0$, the left branch of $r_k(x)$ stays at $1$, while the middle and right branches start to decrease, with the right branch converging to the constant $D$ as $x\ra\infty$. This constant vanishes for $k=k_c$, so that the quantum and non-quantum solutions $r_k$ are the same, while it becomes negative for all $k>k_c$, which implies that the non-quantum $r_k$ is then not the dominant eigenfunction associated with the SCGF, since that function must be positive by the Perron--Frobenius Theorem.

\subsection{Combined solution}

\begin{figure}[t]
\includegraphics{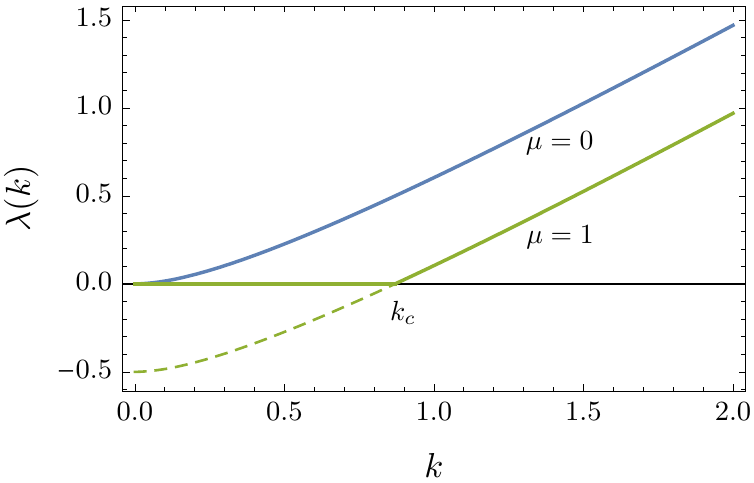}
\caption{(Color online) Scaled cumulant generating function for $\sigma=1$, $a=1$, and different values of $\mu$. The dashed line shows the continuation of the quantum solution as it becomes negative.}
\label{figscgf1}
\end{figure}

The full SCGF is the principal eigenvalue of $\cL_k$ and must therefore be given by the maximum of the two eigenvalues found before:
\be
\lambda(k)= \max\{\lambda_{q}(k),\lambda_{nq}(k)\}=
\left\{
\begin{array}{lll}
\lambda_{nq}(k) & & k\in [0,k_c]\\
\lambda_q(k) & & k>k_c.
\end{array}
\right.
\ee
This result is illustrated in Fig.~\ref{figscgf1} and is consistent with the interpretation of each eigenvalue. On the one hand, the quantum eigenvalue $\lambda_q(k)$ becomes negative for $k<k_c$ and does not converge to $0$ as $k\ra 0$, so that $\lambda(k)$ must be given by the non-quantum eigenvalue $\lambda_{nq}(k)$ satisfying $\lambda_{nq}(0)=0$. On the other hand, for $k>k_c$, the non-quantum solution is no longer valid since part of $r_k(x)$ becomes negative, as noted, which means that $\lambda(k)$ must then be given by the quantum eigenvalue, whose associated eigenfunction is positive and confined. The two eigenvalues cross at $k_c$ (a feature of the non-Hermitian problem), making $\lambda(k)$ continuous, as required by convexity \cite{touchette2009}, but not differentiable at $k_c$. This applies for $\mu>0$. For $\mu=0$, we find $k_c=0$ since $\lambda_{q}(k)\geq 0$, so $\lambda(k)$ is determined only by the quantum solution, which is differentiable.

This result for the SCGF assumes that there are no eigenvalues at the ``top end'' of the spectrum of $\cL_k$ other than the two eigenvalues found before. This is difficult to confirm analytically, due to $\cL_k$ being non-Hermitian, but can be verified indirectly by calculating the rate function associated with this SCGF and by comparing the result with simulation data. There is a subtlety here in that $\lambda(k)$ is non-differentiable at $k_c$, which means that the G\"artner--Ellis Theorem mentioned before does not apply \cite{touchette2009}. However, since the simulation data show that the rate function is convex, we can bypass that theorem to conclude that $I(\rho)$ is also given by the Legendre--Fenchel transform of $\lambda(k)$ \cite{touchette2009}, as expressed in \eqref{eqlf1}. 


\begin{figure}[t]
\includegraphics{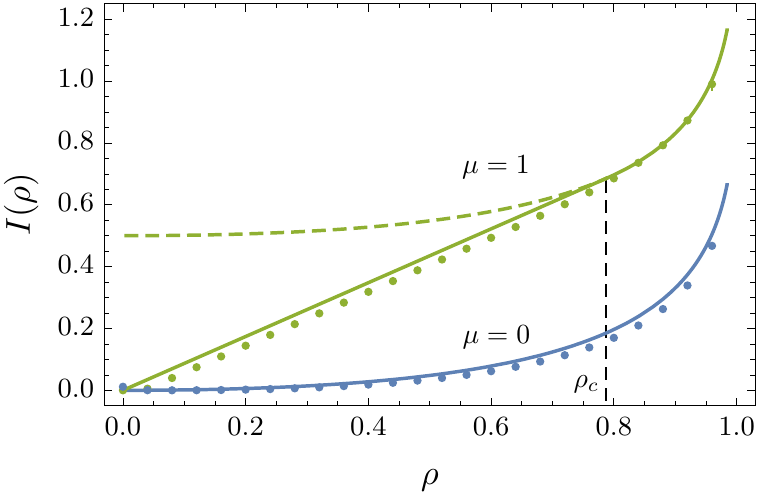}
\caption{(Color online) Rate function for $\sigma=1$, $a=1$, and different values of $\mu$. The dashed line shows the quantum solution, which ceases to be valid below $\rho_c$. The data points are simulation results.} 
\label{figrf1}
\end{figure}

The resulting transform is shown in Fig.~\ref{figrf1}. The main property to notice for $\mu>0$ is that, since $\lambda(k)$ is not differentiable at $k_c$, $I(\rho)$ has a linear branch with slope $k_c$ extending from $\rho=0$, which is the left-slope of $\lambda(k)$ at $k_c$, to a critical occupation $\rho_c$ given by the right slope of $\lambda(k)$ at $k_c$. This follows from known properties of the Legendre--Fenchel transform \cite{touchette2009} and implies that the probability density of $\rho_T$ decays exponentially in both $T$ and $\rho$ according to 
\be
P(\rho_T=\rho)\approx e^{-Tk_c\rho}
\ee 
for $\rho\in[0,\rho_c]$. Above $\rho_c$, $I(\rho)$ is simply the Legendre transform of $\lambda_q(k)$, which is also the rate function obtained for $\mu=0$ shifted by the constant $\mu^2/(2\sigma^2)$ because of Eq.~\eqref{eqqeig1}. In the limit $\mu\ra 0$, both $\rho_c$ and this shift go to $0$, thus recovering the rate function of pure BM given by the Legendre transform of $\lambda_q^0(k)$. This is confirmed by the numerical data, obtained by direct Monte Carlo sampling of the distribution of $\rho_T$ using trajectories of dBM discretized in time with $\Delta t=0.05$ and simulated over $T=30$ for $\mu=0$ and $T=20$ for $\mu=1$ \cite{touchette2011}. In the first case, $10^9$ trajectories were simulated, whereas the second case required $10^{10}$ trajectories to obtain enough statistics for the high occupations.


Naturally, there are no explicit expressions for the SCGF and the rate function, since the former is obtained from a transcendental equation. However, we can easily derive asymptotics for both functions using known asymptotics for the energy levels in the infinite depth limit \cite{barker1991,sprung1992,paul2000,aronstein2000}. For the SCGF, we find
\be
\lambda(k) \approx k+\frac{\pi^2\sigma^3}{2^{5/2}a^3k^{1/2}} -\frac{\pi^2 \sigma^2}{8 a^2}-\frac{\mu^2}{2\sigma^2}
\label{eqscgfapprox1}
\ee  
as $k\ra\infty$, leading by Legendre transform to
\be
I(\rho)\approx\frac{\pi^2 \sigma^2}{8 a^2}+\frac{\mu^2}{2\sigma^2}-\frac{3\pi^{4/3}\sigma^2(1-\rho)^{1/3}}{2^{7/3}a^2}
\ee
as $\rho\ra 1$. This confirms that the probability that dBM stays in $[-a,a]$ for a time $T$ (or, equivalently, that its exit time from $[-a,a]$ is greater than $T$) scales asymptotically as $e^{-TI(1)}$ where
\be
I(1) = \frac{\mu^2}{2\sigma^2}+\frac{\pi^2\sigma^2}{8 a^2}.
\label{eqocc1}
\ee
This case was studied by Pinsky \cite{pinsky1985} (see also Kac \cite{kac1951}). Keeping the first-order term in $\lambda(k)$ also leads to $k_c\approx \mu^2/(2\sigma^2)$ when $\mu\gg 1$, which yields
\be
\rho_c = \lambda'_q(k_c) \approx 1-\frac{\pi^2\sigma^6}{4 a^3 \mu^3}
\ee
when inserted back in Eq.~\eqref{eqscgfapprox1}.


As a side remark, note that it is possible to extrapolate the SCGF for $k<0$ by observing that $I(\rho)$ is defined only for $\rho\in [0,1]$, so we can set $I(\rho)=\infty$ for $\rho\notin [0,1]$, which implies that $\lambda(k)=0$ for all $k<0$. This is not a property of the quantum solution (there is no bound state for $k<0$) nor of the non-quantum solution, but comes rather from the fact that the SCGF is the Legendre--Fenchel transform of the rate function \cite{touchette2009}. With this extension, it can be verified that $\lambda''(k)$ jumps at $k=0$ when $\mu=0$, so one might say that, although pure BM does not have a first-order DPT, it has a second-order DPT at $k=0$. This is a trivial transition, however, that just reflects the fact that $\rho=\lambda'(k)$ starts to grow from $0$ as soon as $k>0$.

\subsection{Driven process}

\begin{figure}[t]
\centering
\includegraphics{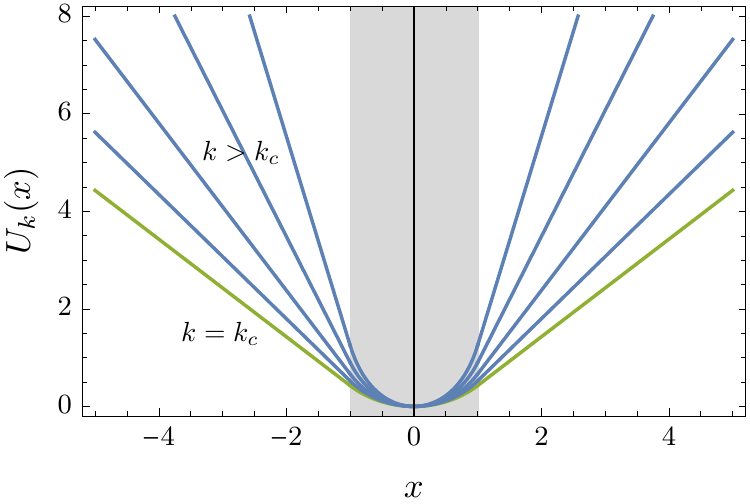}

\vspace*{0.2in}

\hspace*{-0.06in}\includegraphics{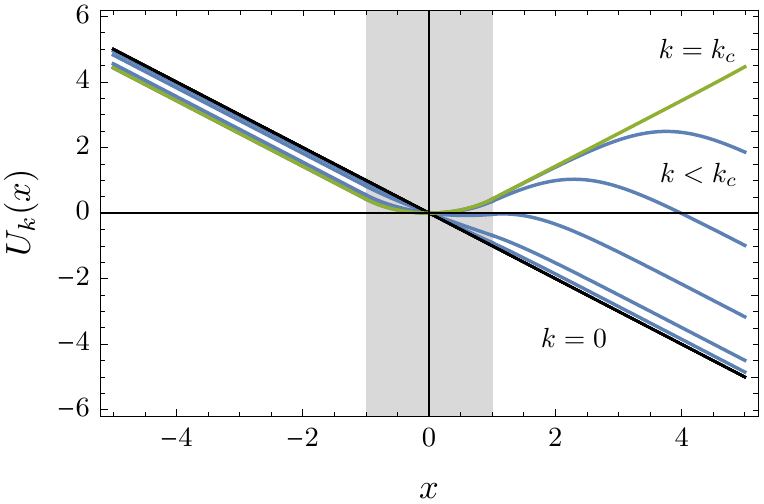}
\caption{(Color online) Top: Effective potential $U_k(x)$ of the driven process in the confinement ($k>k_c)$ regime. Bottom: Effective potential in the escape ($k<k_c)$ regime. Parameters: $\mu=1$, $\sigma=1$, and $a=1$.}
\label{figuk1}
\end{figure}

The non-differentiable point arising in the SCGF for $\mu>0$ signals the appearance of a DPT in the occupation fluctuations, which is first order as $\lambda'(k)$ jumps at $k_c$. To understand the source of this DPT, we plot in Fig.~\ref{figuk1} the effective potential of the driven process,
\be
F_k(x) = -U'_k(x),
\label{eqf1}
\ee 
associated with its modified force or drift. From the expression \eqref{eqeff1} of this drift, we thus find
\be
U_k(x) = -\mu x - \sigma^2 \ln r_k(x) + c
\label{equk1}
\ee
or
\be
U_k(x)=-\sigma^2 \ln \psi_k(x) +c
\label{equk2}
\ee
when $r_k$ is given by the quantum solution \eqref{eqqsol1}, following the symmetrization of Eq.~\eqref{eqtransf1}. In both cases, $c$ is an integration constant set such that $U_k(0)=0$.


The top plot in Fig.~\ref{figuk1} shows the latter potential obtained for values $k\geq k_c$, which characterize the occupation fluctuations $\rho\in [\rho_c,1]$ above the critical occupation $\rho_c$. In this case, we see that $U_k(x)$ is a confining potential, which means that those occupations are effectively created by a driven process that is ergodic and thus confined with stationary density $p_k(x)=\psi_k(x)^2$ such that 
\be
\rho = \int_{-a}^a p_k(x)\, dx = \lambda'(k).
\ee
For this occupation value, it can be shown \cite{chetrite2015} that the rate function is given by
\be
I(\rho)=\frac{1}{2\sigma^2}\int_{-\infty}^\infty (F_k(x)-\mu)^2 p_k(x)\, dx
\ee
which reduces to
\be
I(\rho) = \frac{\mu^2}{2\sigma^2}+\frac{\sigma^2}{2}\int_{-\infty}^\infty \psi_k'(x)^2\, dx
\label{eqrf2}
\ee
from \eqref{equk2}, \eqref{eqf1}, and the fact that $\psi_k(x)$ is even. In the limit $k\ra \infty$, the two linear branches of $U_k(x)$ becomes infinitely steep, creating two logarithmic singularities close to the boundaries of $[-a,a]$ \cite{pinsky1985}. This follows because the quantum well then becomes infinite, so the ground-state wavefunction $\psi_k(x)$ converges to
\be
\psi_\infty(x) =\frac{1}{\sqrt{a}}\cos \left(\frac{\pi x}{2a}\right)
\label{eqpsiq1}
\ee
for $x\in [-a,a]$ and $\psi_\infty(x)=0$ otherwise. The corresponding density $p_\infty(x)$ is then all supported in $[-a,a]$, leading to $\rho=1$. Inserting \eqref{eqpsiq1} in \eqref{eqrf2} also confirms the result in \eqref{eqocc1} for the probability that dBM stays in $[-a,a]$.


It should be noted that the driven process in this confinement regime ($k\geq k_c$) does not depend on $\mu$, since $\psi_k$ itself does not depend on that parameter, as noted before. This remarkable property can be understood by noting that atypical paths of dBM that stay for a very long time in the occupation interval must not have a drift that would otherwise take them outside the interval. In other words, the noise must ``cancel'', so to speak, the drift of $X_t$ for the process to stay in $[-a,a]$. The likelihood of this happening does depend on $\mu$, however, and explains why the SCGF and the rate function depend on $\mu$. In fact, we know from \eqref{eqqeig1} that
\be
\lambda(k) = \left.\lambda(k)\right|_{\mu=0}-\frac{\mu^2}{2\sigma^2}
\ee
for $k\geq k_c$, leading to
\be
I(\rho) = \left.I(\rho)\right|_{\mu=0}+\frac{\mu^2}{2\sigma^2},
\ee
for $\rho\geq \rho_c$. The latter result can also be derived from Girsanov's Theorem by noting that the Radon--Nikodym derivative of the driven process with respect to dBM has an extra $\mu^2/(2\sigma^2)$ compared to BM \cite{morters2010}. Alternatively, we can notice that the integral in \eqref{eqrf2} is nothing but the level-2 rate function of the BM \cite{chetrite2015}, expressed in terms of $\psi_k=\sqrt{p_k}$. 

The behavior of the driven process in the complementary regime where $k\in [0,k_c)$ is very different. There we see from the bottom plot of Fig.~\ref{figuk1} that the effective potential $U_k(x)$ is not confining, which implies that the driven process escapes $[-a,a]$ as $T\ra\infty$, leading to $\rho_T\ra 0$ with probability 1 in that limit, as for dBM itself ($k=0$). This is source of the DPT: as $k$ is varied across $k_c$, the driven process changes abruptly from being deconfined to confined, with its typical occupation jumping from $0$ to $\rho_c$. This means physically that we have an abrupt change, as a function of $k$, in the process or mechanism responsible for the occupation fluctuations, which is what a dynamical phase transition is. 

It is important to note that this transition does not appear at the level of the rate function because the latter is expressed as a function of the occupation fraction, which can be fixed by conditioning to any value in $[0,1]$, including any value between $0$ and $\rho_c$. The only property of the DPT reflected in $I(\rho)$ is the linear branch interpolating between $0$ and $\rho_c$. This is similar to first-order phase transitions that appear at equilibrium as a function of temperature and that lead to ``phase co-existence'' or ``phase mixture'' lines in the entropy (e.g., the liquid-vapour phase transition of water as a function of temperature leading to a phase co-existence in density) \cite{reichl1980}. 

\begin{figure}[t]
\includegraphics{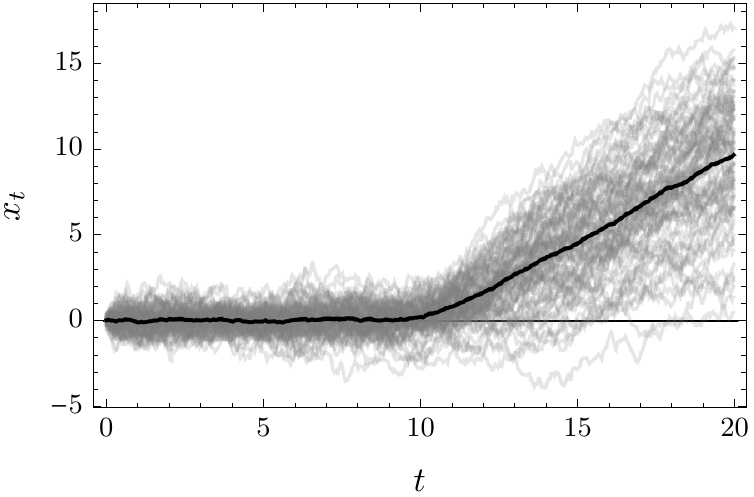}
\caption{Paths of dBM conditioned to stay in $[-1,1]$ for approximately half of the time ($\rho=0.5\pm 0.01$). Parameters: $\mu=1$, $\sigma=1$, $T=20$, $dt=0.05$. Grey lines: 86 paths obtained out of $10^6$ simulated sample paths. Black line: Average of the paths showing that they escape after some time with an average speed $\mu=1$. Similar results are found for any $\rho\in (0,\rho_c)$.}
\label{figinst1}
\end{figure}

In our case, the linear branch is found from simulations to be created by paths having two ``coexisting'' parts or periods, as shown in Fig.~\ref{figinst1}, where the process is first confined according to the driven process with occupation $\rho_c$ for a fraction $\alpha\in [0,1]$ of the total time, approximately given by $\alpha=\rho/\rho_c$, before escaping like normal dBM for the remaining time. The occupation fraction realized by these paths is thus
\be
\rho_T = \rho_c \frac{\alpha T}{T}+0\frac{(1-\alpha)T}{T}=\rho.
\ee
Moreover, since the occupation is additive in time, its probability density factorizes over each period: 
\be
P(\rho_T=\rho) \approx e^{-\alpha T I(\rho_c)} e^{-(1-\alpha)T I(0)},
\ee
leading to
\be
I(\rho) = \alpha I(\rho_c)=\frac{\rho}{\rho_c} I(\rho_c).
\ee

While this predicts the correct rate function, it is important to note that the ``coexistence'' region is not described completely by the driven process, since there is no link between $k$ and $\rho$ via Eq.~\eqref{eqlf2} whenever $\lambda'(k)$ is not differentiable, that is, whenever there is a first-order DPT. Our model shows, in fact, a known case of nonequivalence of ensembles for Markov processes referred to as ``partial equivalence'' \cite{touchette2015,szavits2015}. This is clear also by noting that, since the driven process is homogeneous and ergodic, it cannot describe a non-homogeneous process that has two different parts -- confined and deconfined. The argument above only captures the confined part where the process conditioned on reaching an occupation $\rho_T\in (0,\rho_c)$ mimics the driven process with occupation $\rho_c$ for \emph{some} of the time before escaping like a normal dBM.


The escape itself is analogous to Markov processes with absorbing states, which provide the simplest examples of rate functions having linear parts (see \cite{whitelam2018} and the appendix of \cite{coghi2018b}). The absorbing state in our case is represented by the complement of $[-a,a]$ (i.e., the state ``$ X_t\notin[-a,a]$''), which is eventually reached by dBM and serves as a trap for it, as this process is not recurrent, meaning that it has a zero probability to ever return close to $x=0$ \cite{borodin2015}. We then say that the process is transient. The standard BM is recurrent and does not have a DPT. We will come back to this point in the next sections. 

This analogy with absorbing Markov processes is qualitatively correct, but cannot be used to predict the value of the critical density $\rho_c$ because the ``occupation'' process $Y_t$ obtained by ``coarse-graining'' dBM into the state $Y_t=1$ when $X_t\in [-a,a]$ and $Y_t=0$ otherwise is not Markovian. In fact, this transformation has the form of a hidden Markov process \cite{ephraim2002} in which $X_t$ is the hidden layer and $Y_t$ the visible layer. The latter process appears also \emph{not} to be semi-Markovian, as $X_t$ can go in and out of $[-a,a]$ in many ways (e.g., in and out from $a$ versus in from $a$ but out from $-a$), which are not equivalent or symmetric when there is a drift. 

\section{Confined model}

\begin{figure*}[t]
\centering
\includegraphics[width=0.32\textwidth]{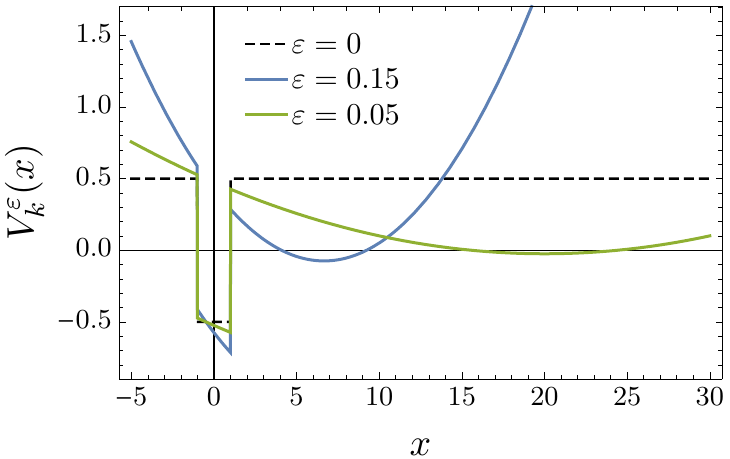}%
\hspace*{0.1in}%
\includegraphics[width=0.32\textwidth]{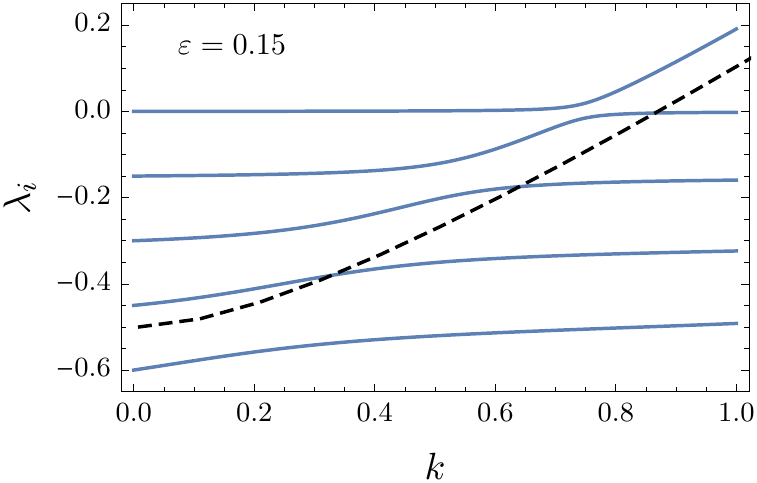}%
\hspace*{0.1in}%
\includegraphics[width=0.32\textwidth]{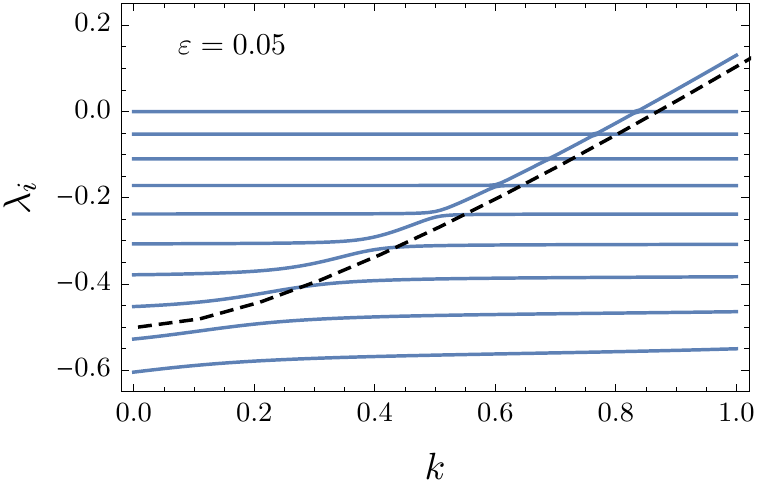}
\caption{(Color online) Left: Large deviation potential of the confined model for $\mu=1$, $\sigma=1$, $a=1$, and different confinement values $\eps$. The potential depth is $k=1$. Middle and right: Top part of the spectrum. The dashed line shows the eigenvalue of the quantum solution for $\eps=0$.}
\label{figcpot1}
\end{figure*}

To confirm the results of the previous section, we consider a variant of the model satisfying the SDE,
\be
dX_t = (\mu-\eps X_t) dt+\sigma dW_t,
\ee
which is an Ornstein-Uhlenbeck process evolving in the quadratic or harmonic potential
\be
U^\eps(x) = -\mu x+\frac{\eps x^2}{2}
\ee
with $\eps>0$ playing the role of a friction parameter. For $\eps=0$, we recover dBM. 

The reason for considering this model is that $X_t$ is now ergodic, since $U^\eps(x)$ is confining, so we expect no DPT to occur in the occupation large deviations. This is confirmed by noting that the tilted generator $\cL_k$ in this case cannot have any crossing eigenvalues as a function of $k$, since $L$ and therefore $\cL_k$ are conjugated to a Sturm--Liouville problem \cite{pavliotis2014}, corresponding to the quantum problem obtained by the symmetrization \eqref{eqsym1}. By considering the deconfinement limit $\eps\ra 0$, we want to understand how a crossing of eigenvalues can occur in the quantum problem, similarly to quantum phase transitions, and how this gives rise to the large deviation DPT. 

\begin{figure}[t]
\centering
\includegraphics{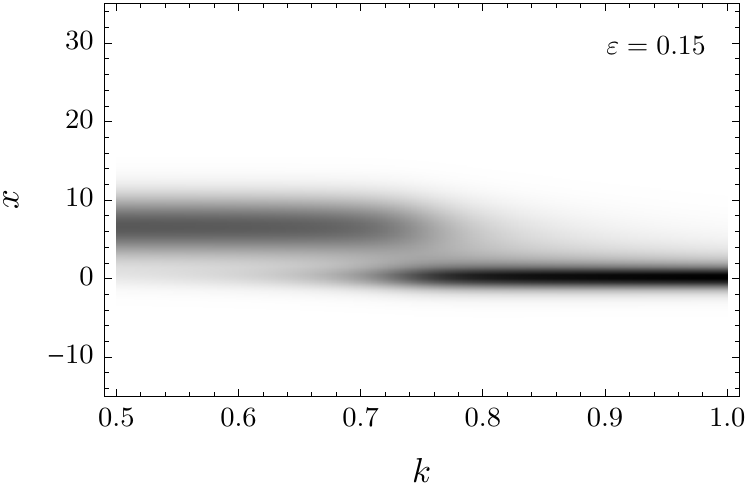}

\vspace*{0.2in}

\includegraphics{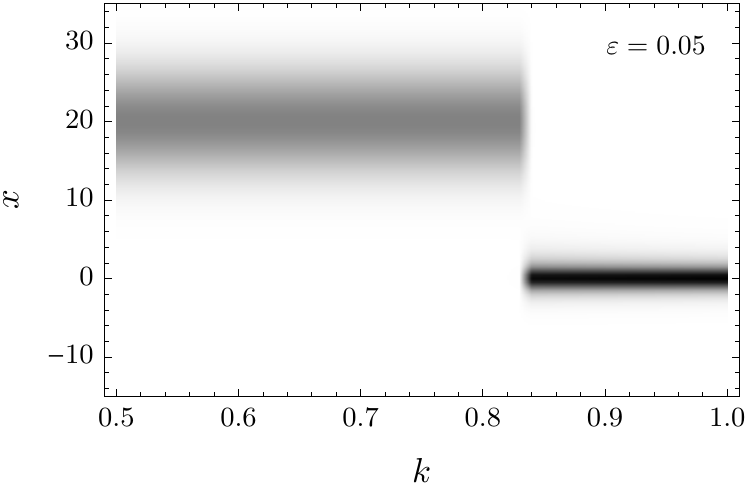}
\caption{Density plot of the ground-state wavefunction $\psi_k(x)$ associated with the confined potential $V_k^\eps(x)$. Parameters: $\mu=1$, $\sigma=1$, and $a=1$.}
\label{figpsik1}
\end{figure}

The same analysis could be done in principle with other types of confinements or compactifications, for example, by considering dBM in a finite box with reflecting or periodic boundary conditions \footnote{Note that it is not enough to confine the quantum problem associated with the large deviations; we must confine the original stochastic motion.}. However, the parabolic confinement above is simpler to deal with, as the quantum potential associated with the large deviations of $\rho_T$ \cite{touchette2017} is a parabola
\be
V^\eps_k(x)= \frac{\eps^2(x-\mu/\eps)^2}{2\sigma^2}-\frac{\eps}{2} -k\idf_{[-a,a]}(x),
\ee
punctured by the indicator function, which creates a well of depth $-k$, as shown in Fig.~\ref{figcpot1}. The spectrum associated with this potential can easily be calculated numerically using mesh methods \footnote{We use the Mathematica functions NDEigenvalues and NDEigensystem with an adaptive mesh method over a spatial range large enough to impose decaying boundary conditions for the wavefunction $\psi_k$.}. The results are presented in Fig.~\ref{figcpot1}, with a minus sign to account for the connection with the SCGF, for two values of the confinement parameter, namely, $\eps=0.15$ and $\eps=0.05$. In the first case, the dominant eigenvalue is close to $0$ as $k$ increases and then starts to grow after some $k$ close to the critical value $k_c$ found in the previous section. This reflects the fact that the ground-state energy of $V_k^\eps(x)$ is close to zero when the well is shallow, the ground state $\psi_k(x)$ being localized in the minimum $x^*=\mu/\eps$ of the potential, and starts to decrease when the well becomes deep enough, with $\psi_k(x)$ then transitioning in the well, as illustrated in Fig.~\ref{figpsik1}. 

This transition is a smooth crossover that becomes discontinuous in the limit where $\eps\ra 0$, as can be seen from the results obtained for $\eps=0.05$. When $\eps>0$, all eigenvalues vary continuously in $k$ and have avoided crossings, reaching plateaus close to the eigenvalues of the harmonic well, except for the dominant one. As $\eps\ra 0$, the avoided crossings get closer and effectively become, for the two largest eigenvalues, a crossing between the eigenvalue $\lambda=0$ and a positive eigenvalue that aligns itself on the quantum solution found in the previous section (dashed line in Fig.~\ref{figcpot1}), thereby confirming the results of that section. The first-order transition between the escape and confined regimes of the driven process is also seen in the ground-state wavefunction $\psi_k$, shown in Fig.~\ref{figpsik1}, which jumps from being localized around $x^*$, and so escapes to infinity as $\eps\ra 0$, to being localized and confined in the well as soon as its depth reaches the critical value $k_c$.

This transition, which is effectively a large deviation analog of a ground-state quantum phase transition \cite{sachdev2011}, shows up remarkably in all the other eigenvalues above $-\mu^2/(2\sigma^2)$, which get closer as the harmonic potential opens up in the limit $\eps\ra 0$, thus creating an infinite number of ``effective'' crossings that line up on the quantum eigenvalue of the square well. This is an interesting phenomenon, showing that the simple, confined model not only has a \emph{ground-state} quantum phase transition, but also an infinite number of \emph{excited-state} quantum phase transitions \cite{caprio2008}, which can be studied analytically using a delta perturbation of the harmonic well \cite{janev1974,busch1998,viana2011,gadella2011}.

To close this analysis, note that there is no DPT for $\mu=0$ because the ground-state eigenvalue decreases in a regular way with $k$ when the well is exactly in the middle of the harmonic well. This confirms overall that the DPT appears rigorously only when $X_t$ is \emph{transient}, that is, when it is not ergodic ($\eps=0$), and there is a drift. In spite of this, we see from the results obtained for $\eps=0.05$ that the process can be ergodic and still show all the signs of a DPT if it is weakly confined. This is similar to equilibrium phase transitions, which are defined mathematically in the thermodynamic limit, but which nevertheless appear in macroscopic systems that have a large yet finite volume.

This is an important point, as all simulations or measurements of large deviations are performed ultimately in finite time and on finite-size systems. To say that $X_t$ is transient involves an infinite-time limit, as does the definition of large deviation functions, which must be compared with basic timescales of the process considered. In our case, the onset of the DPT should be observed in any confined version of the dBM if the well width $a$ is much smaller than the confinement length-scale $\ell$, given here by $\ell\sim x^*=\mu/\eps$. In other words, the DPT should be seen whenever the timescale $\tau_\esc=a/\mu$ needed to escape the well is much smaller than the timescale $\tau_\conf=\ell/\mu\sim 1/\eps$ needed for dBM to reach the boundaries, and thus to ``feel'' the effect of confinement \footnote{The timescale $1/\eps$ is also the relaxation time to the stationary distribution.}. 

This is only a rough estimate, as the noise power $\sigma$, acting as a kind of temperature, also ``rounds'' the DPT whenever $\eps>0$. Moreover, the convergence in time of the large deviation functions will be influenced by these timescales. However, the basic point remains that the DPT should be seen whenever $\tau_\conf$ is large enough that the confined motion behaves as a real transient dBM that has very small probability of returning to the occupation interval over the time $T$ used to calculate the large deviations.

\section{Conclusion}

We have shown that DPTs can arise in the large deviations of systems as simple as dBM. This transition is different from other DPTs reported in recent years, as it does not involve a thermodynamic, macroscopic or low-noise limit, in addition to the large-time limit. In our case, the DPT is a transition between two regimes of long-time fluctuations: a localized regime, where the process behaves in a confined and ergodic way to realize high occupations, and a delocalized regime, where it escapes away from the origin, similarly to Markov processes with absorbing states, due to the drift to realize low occupations. The transition is discontinuous (first-order) and arises because the motion is transient, although its effect can also be seen when the motion is weakly confined. For this reason, we expect it to be observable in physical systems even if they can only be probed on finite space and time scales.

The connection with the transient property and quantum potentials suggest other processes that should have the same occupation DPT, in particular, BM in $d\geq 3$ dimensions or its radial projection, the Bessel process, which are transient without drift \cite{borodin2015}. This is confirmed by noting that the quantum well (spherical or hypercube) in $d\geq 3$ dimensions has a ground state only below a critical depth. The precise form of the SCGF in this case is however unknown: it should be non-trivial since this function is convex in $k$ and so must be continuous despite the discontinuity in the ground state energy. For $d=2$, there is no first-order DPT, since BM is then recurrent \footnote{BM is actually not recurrent for $d=2$ but \emph{neighborhood recurrent}: the probability that it returns to the origin is $0$, but the probability that it returns in a neighborhood of the origin (or any point in the plane) is $1$ \cite{morters2010}.}, although there could be a second-order DPT coming from a weakly bound ground state. This can be verified in principle from known estimates of occupation times of BM in two- and three-dimensional balls \cite{chan1994,gruet1996}.

The square well with one infinite wall provides another example related to reflected BM for which there is a bound state only below a certain depth. In this case, however, Neumann instead of Dirichlet boundary conditions must be used to enforce reflection on the wall, which leads the well to have a bound state for any depth, so there is no DPT. This is consistent with the fact that reflected BM is recurrent \cite{borodin2015}. Note that there is also no DPT for dBM if we take the occupation interval to be the half-line $[0,\infty)$ leading to the arcsine law \cite{levy1940,akahori1995,dassios1995}.

In principle, occupation DPTs could also arise in ergodic and therefore recurrent processes that becomes transient upon conditioning on their large deviations, showing that it is not the transient property as such that is important for the transition to occur, but the \emph{possibility} for the process to be transient. As important, mathematically, is the non-Hermitian nature of the tilted generator underlying the large deviations, since crossings of eigenvalues are precluded in general for Schr\"odinger-like Hermitian operators with confining potentials. Proving rigorous results in that direction is a challenging problem, however, as there are very few general results known about the spectrum of non-Hermitian operators on non-compact spaces.

The way that the DPT appears can be further related to wetting transitions in absorption phenomena \cite{kroll1983}, magnetic depinning transitions in superconductors \cite{hatano1996,hatano1997,feinberg1999}, population dynamics in inhomogeneous environments \cite{nelson1998,dahmen2000}, and, superficially, to biased random walks in random environments \cite{mehra2002}. In all these cases, a localization transition occurs when the potential created by a substrate surface, impurities or spatial inhomogeneities becomes attractive enough. In the case of superconductors, one can even map the dBM model exactly to a non-Hermitian quantum model studied by Hatano and Nelson \cite{hatano1997} in which $x$ represents the coordinate of a magnetic flux line perpendicular to a pinning defect, $t$ is the coordinate parallel to the defect, while $\mu$ is proportional to the magnetic field generated by the superconducting current. The localization regime of this model was studied similarly to here using the symmetrization \eqref{eqsym1}, which is referred to as an imaginary gauge transformation, with results similar to ours (see, in particular, Sec.~IV of \cite{hatano1997}). The physical interpretation of the results, of course, is completely different. 

To finish, we want to mention that the occupation DPT of dBM will also arise in geometric BM, since the latter is simply an exponential transformation of dBM. This opens up the study of large deviations and DPTs in the context of finance, where geometric BM is used as a basic model of stock prices while occupation conditioning is related to the pricing of options \cite{bouchaud2000b,akahori1995,dassios1995}, holding periods, and the ``survival'' of equities~\cite{brown1995}. 

\begin{acknowledgments}
H.T. thanks Michael Kastner for computer access, Jens Uwe N\"ockel for useful Mathematica code, as well as Rapha\"el Chetrite, Peter Grassberger, Baruch Meerson, David Mukamel, Paul Krapivsky, and especially Yariv Kafri for useful discussions. P.T.\ was supported by a DAAD PhD Scholarship administered by AIMS. H.T.\ was supported by NRF South Africa (Grants No.\ 90322 and No.\ 96199).
\end{acknowledgments}

\bibliography{masterbib}

\end{document}